\input{aipcheck}

\documentclass[
  ,draft            
  ]
  {aipproc}

\layoutstyle{6x9}

\begin{document}

\title{Bose-Einstein correlations from "within"}



\author{O.V. Utyuzh}{
           address={The Andrzej So\l tan Institute for Nuclear Studies;
           Ho\.za 69; 00-681 Warsaw, Poland }
           }
\author{G. Wilk}{
           address={The Andrzej So\l tan Institute for Nuclear Studies;
           Ho\.za 69; 00-681 Warsaw, Poland }
           }
\author{Z.W\l odarczyk}{
  address={Institute of Physics, \'Swi\c{e}tokrzyska Academy,
         \'Swi\c{e}tokrzyska 15; 25-406 Kielce, Poland}
         }

\begin{abstract}
We describe an attempt to model numerically Bose-Einstein correlations
(BEC) from "within", i.e., by using them as the most fundamental
ingredient of some Monte Carlo event generator (MC) rather than
considering them as a kind of (more or less important, depending on the
actual situation) "afterburner", which inevitably changes original
physical content of the MC code used to model multiparticle production
process.
\end{abstract}

\maketitle



\paragraph{Introduction}

The problem of BEC is so well known that we shall skip introductory
remarks (referring in that matter to other presentations at this
conference and to \cite{QCM} for more information) and instead we shall
start right away with our subject concerning {\it the proper numerical
modelling of BEC}, which we call {\it BEC from within}. Although
phenomenon of BEC is with us from the very beginning of the systematic
investigation of multiparticle production processes, its modelling is
still virtually nonexisting. With the exception of attempts presented in
\cite{ZAJC,OMT} all other approaches are tacitly assuming that on the
whole BEC constitute only a small effect and it is therefore justify to
add it in some way to the already known outputs of the MC event
generators widely used to model results of high energy collisions (in the
form of the so called {\it afterburner}) \footnote{Methods presented in
\cite{ZAJC} were never used in practice, only \cite{OMT} applied their
approach to $e^+e^-$ annihilation data.}. There are two types of such
afterburners:
\begin{itemize}
\item[$(a)$] those modifying accordingly energy-momenta of identical
secondaries (and correcting afterwards the whole sample for
energy-momentum conservation) - they apply to each event separately;
\item[$(b)$] those selecting events which already have (due to some
fluctuation present in any MC code) right energy-momenta of identical
secondaries and counting them (by introducing some weights) many times
(see \cite{QCM} for references) - in this case energy-momentum balance is
left intact but instead the particle spectra provided by MC code are
distorted (albeit in most cases only slightly); they apply only to all
events.
\end{itemize}

\paragraph{Example of modifications introduced by afterburner}

\begin{figure}
  \includegraphics[height=.195\textheight]{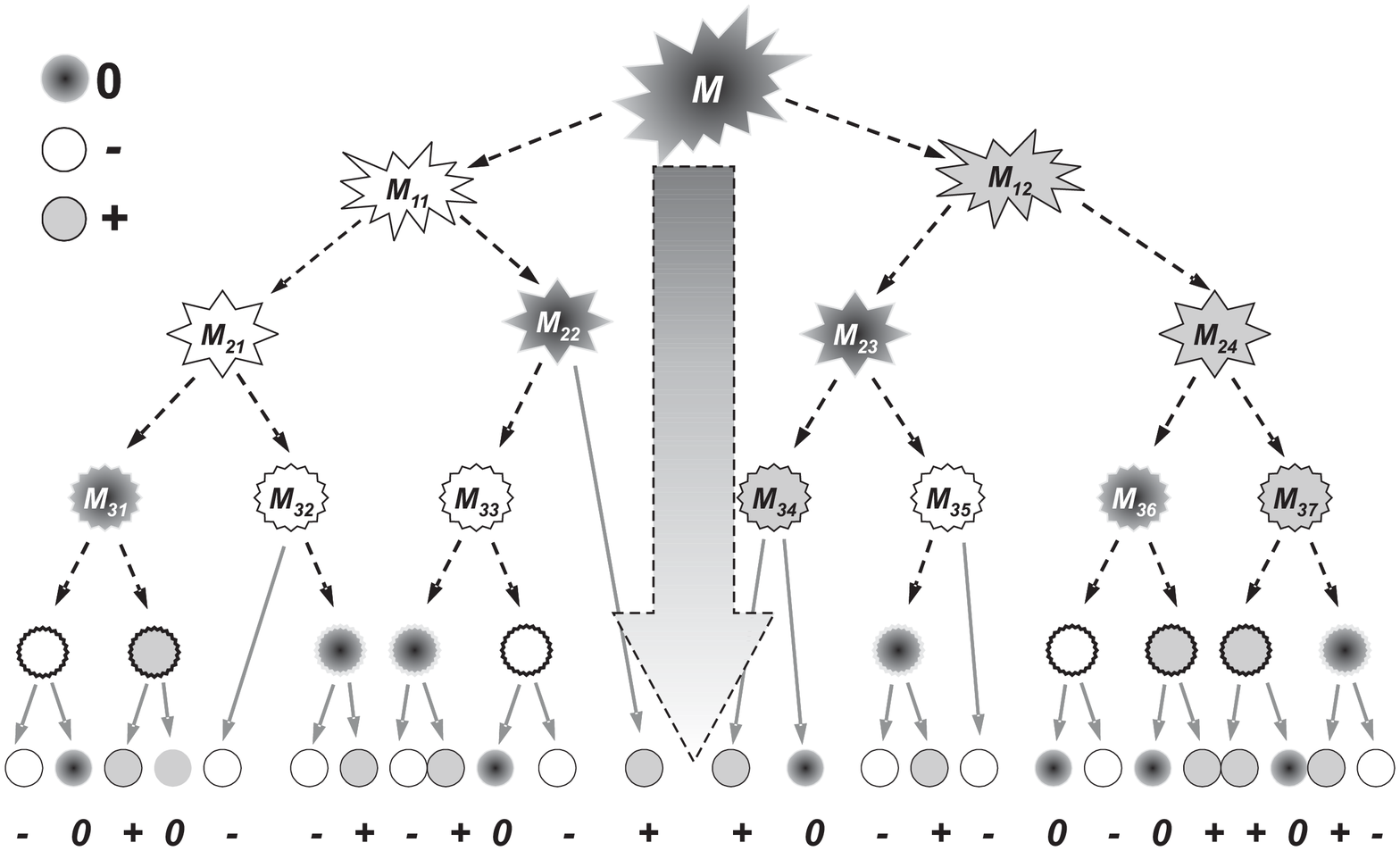}
  \hspace{3mm}
    \includegraphics[height=.195\textheight]{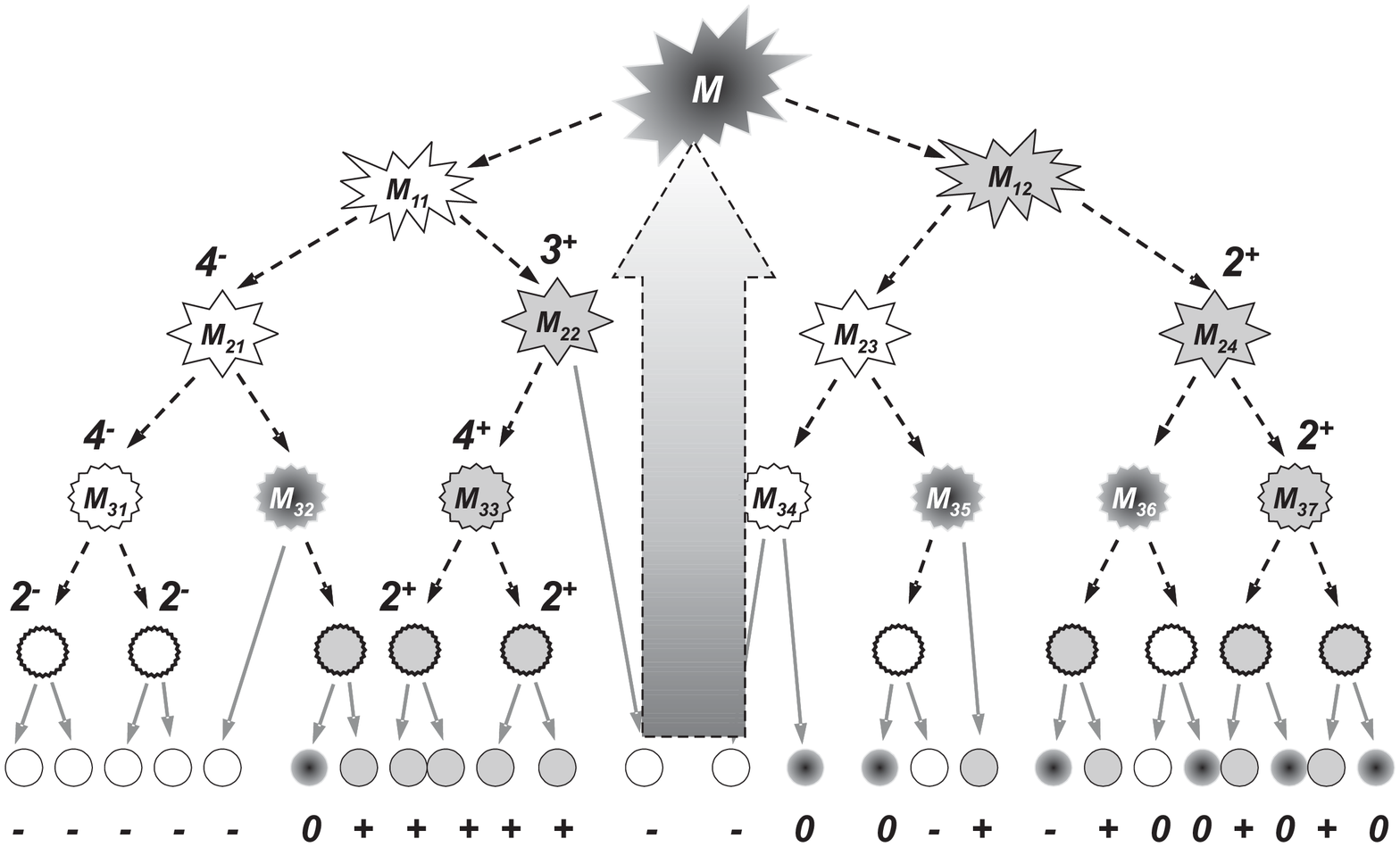}
  \caption{Example of charge flows in MC code using simple cascade model
           for hadronization \protect\cite{CAS}: left panel - no effect of BEC
           observed; right panel - after applying afterburner described in
           \protect\cite{UWW} (based on new assignement of charges to the produced
           particles) one has BEC present at the cost of appearance
           of multicharged vertices.}
  \label{fig:Fig1}
\end{figure}

It must be stressed that modifications that such afterburners introduce
to {\it physical background} behind a given MC code were never
investigated. Tacit assumption made is that they are small and therefore
irrelevant \footnote{Actually the frequent practice is to use the
original MC data to obtain single particle spectra and to calculate the
corresponding BEC {\it afterwards} by means of some afterburner.}. The
problem is that with the MC codes in use at present it is practically
impossible to estimate the nature and strength of such changes. In can be
done only by using some simple scheme of cascading, for example simple
cascade model for hadronization developed by us some time ago \cite{CAS}.
This model {\it per se} leads to no BEC (cf. left panel of Fig.
\ref{fig:Fig1}). To obtain effect of BEC one has to add to it some simple
afterburner, for example the one proposed by us in \cite{UWW}. In it one
preserves the whole space-time and energy-momentum structure of each
event but changes the charge assignment to secondaries in such way that
clusters of identically charged particles occur (the original number of
$(+)$, $(-)$ and neutrals remains the same). This automatically leads to
BEC \footnote{Notice that such reassignment of charges results in the
effective bunching of momenta very much similar to that assumed in the
method $(a)$ mentioned above \cite{QCM}.}. The right panel of Fig.
\ref{fig:Fig1} shows what are the changes introduced by this afterburner
in the original cascade: {\it the multicharged vertices} occur now (with
total charge in the whole event remaining conserved). It means that, {\it
in principle}, one could obtain BEC directly from MC (i.e., without any
afterburner) allowing in it for the appearance of such vertices
(according to some prescribed scheme - for example as some multicharged
clusters, such possibility has been already mentioned in \cite{BUSH})).
However, it would be then extremely difficult to run such cascade till
the very end without producing spurious multicharged particles not
observed in nature. Our afterburner can be then considered as a kind of
shortcut realization of such possibility with well defined physical
consequences. We argue therefore that each afterburner changes original
MC code it is attached to in a similar fashion but this statement cannot
be at the moment substantiated\footnote{Notice that the problem, which is
clearly visible in the CAS model, is not at all straightforward in other
approaches. However, at least in the string-type models of hadronization,
one can imagine that it could proceed through the formation of charged
(instead of neutral) color dipoles, i.e., by allowing formation of
multi(like)charged systems of opposite signs out of vacuum when breaking
the string. Because only a tiny fraction of such processes seems to be
enough in getting BEC in the case of CAS model, it would probably be
quite acceptable modification. It is worth to mention at this point that
there is also another possibility in such models, namely when strings are
nearby in the phase space one can imagine that production of given charge
with one string enhances emission of the same charge from the string
nearby - in this case one would have a kind of {\it stimulated emission}
discussed already in \cite{GN}.}.

\paragraph{BEC from "within"}

The above observation was one of our motivations to look for the MC
scheme, which would be build around BEC rather than starting from some
single-particle observables. This idea has been for the first time used
in \cite{OMT} where particles were selected from a grand-canonical-like
distribution with temperature $T$ and chemical potential $\mu$ chosen in
such way as to describe rapidity and multiparticle
distributions\footnote{Actually, in \cite{OMT}, which was using
information theory approach based on Shannon entropy, $T$ and $\mu$ are,
in fact, two lagrange multipliers obtained from energy conservation and
charge conservation constraints.} and particles were then distributed
into some rapidity cells of given width, each cell containing only
particles of the same charge; the method was then (quite successfully)
used to describe $e^+e^-$ annihilation data. Introduction of these cells
was the real origin of the BEC observed. Actually the idea that BEC
demands that particles of the same charge are emitted from some cells
(named {\it elementary emitting cells} - EEC) was proposed earlier in
\cite{BSWW} and consist also cornerstone of our present proposal, which
can be regarded as generalization of that presented in \cite{OMT}. This
is nothing else but attempt to numerically realize the bunching of
particles as quantum statistical effect used already in connection with
BEC long time ago \cite{GN} and is done by dividing all available energy
among particles (taken here as being pions) in such way that a number
$n_{cell}$ of EEC's, each containing particles of the same charge, is
formed, with multiplicity of particles in each EEC followinig
Bose-Einstein (or geometrical) distribution\footnote{To this end to the
first particle selected in a given EEC one adds (up to first failure
after which new EEC is selected) another particles with probability
$P=P_0\cdot e^{-E/T}$, where $P_0\in (0,1)$ is constant, $E$ is energy of
the first particle and $T$ parameter (corresponding to temperature in
thermal models). Such form of $P$ ensures the characteristic
Bose-Einstein form of energy distribution.}. When energy distribution
from which particles are selected is thermal-like then $P(n_{cell})$ is
Poissonian and the total multiplicity distribution is of P\`olya-Aeppli
type \cite{PA}, closely resembling Negative Binomial distribution
obtained in the so called {\it clan model} \cite{NBD} (which differs only
by the fact that particles in clans are distributed according to
logarithmic distribution, not geometrical one \cite{QCM}). Particles in a
given EEC can have energies spread around the energy $E_1$ of the first
particle defining this EEC with some width $\sigma$. With such energy
spreading allowed one gets quite reasonable results for $C_2(Q_{inv})$
distributions (see \cite{QCM} for details)\footnote{Actually, as was
shown in \cite{Kozlov} (see also \cite{Zal}), this spreading is crucial
to obtain the proper shape of $C_2$ function.}.

Here we would like to present extension to this algorithm, which in
addition to bunching accounts also for the symmetrization of the
two-particle wave function (not used before) and allows to obtain in
addition to $C_2(Q_{inv})$ also $C_2(Q_{x,y,z})$, i.e., in a sense it is
$3$-dimensional extension of our algorithm. This extension is based on
the observation that symmetrization correlates the energy-momenta of
particles with their space-time locations. The bunching of particles
considered before was done only in the energy-momentum space and left us
with a number of EEC's, each with a number of particles with well defined
energies, $E_i$ (and momenta $p_i = (E_i^2 - m^2)^{1/2}$). Each EEC is
build around some "seed" particle, which is taken as particle $i=1$. This
was enough to get $C_2(Q_{inv})$ (see \cite{QCM}), but not
$C_2(Q_{x,y,z})$ involving components of $p_i$, $p_{i(x,y,z)}$. To assign
them one has first choose some space-time positions for particles in a
given EEC taking them from some distribution function $\rho(r,t)$.
Actually in what follows we shall use only {\it static source
approximation}, i.e., hadronization is instantaneously and therefore
$\rho(r,t) = \rho(r)$.  Now $p_{i(x,y,z)}$ have to be correlated with the
corresponding space positions, $r_i=(x_i,y_i,z_i)$, in the way emerging
from the symmetrization of the wave functions resulting (for the plane
wave approximation) in the famous $1+\cos(\delta p\cdot\delta r)$
expression. Technically this is achieved by accepting only such momenta
$p_{i(x,y,z)}$, which for given $r_i$ lead to $\cos(\delta p\cdot\delta
r) \le 2\cdot Rand -1$ where $Rand$ is random number uniformly chosen
from interval $(0,1)$.

In Fig. \ref{fig:Fig2} we present examples of our new results (the elder
results can be found in \cite{QCM}) obtained with full, $3$-dimensional
version of our model for $\rho(r)$ being sphere of radius $R=1$ fm and
assuming that all $p_{i(x,y,z)}$ are spherically symmetric. As one can
see now in addition to $C_2(Q_{inv})$ one has also corresponding to it
$p_{i(x,y,z)}$ $C_2(Q_{x,y,z})$.

\begin{figure}
  \includegraphics[height=.25\textheight]{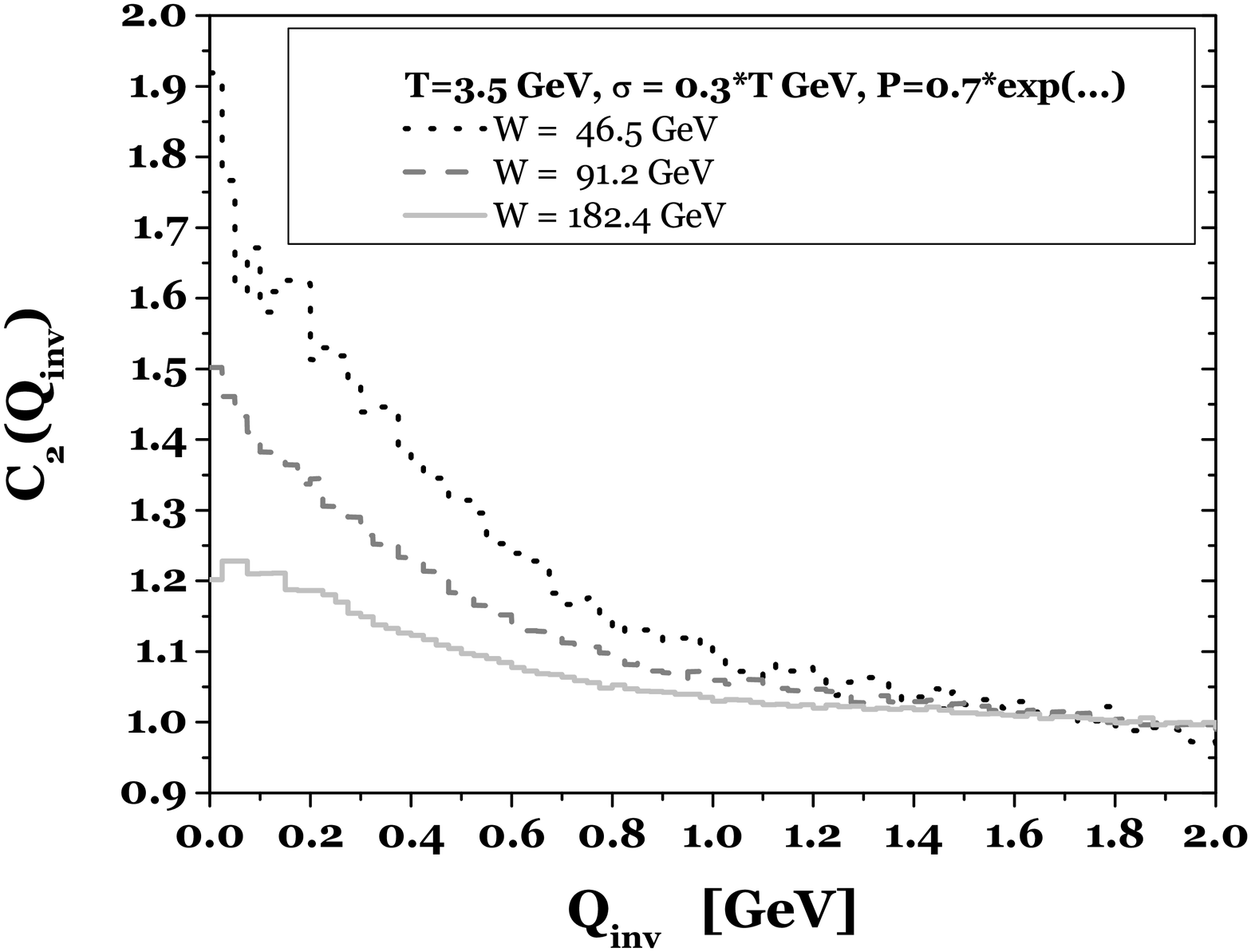}
  \hspace{3mm}
    \includegraphics[height=.25\textheight]{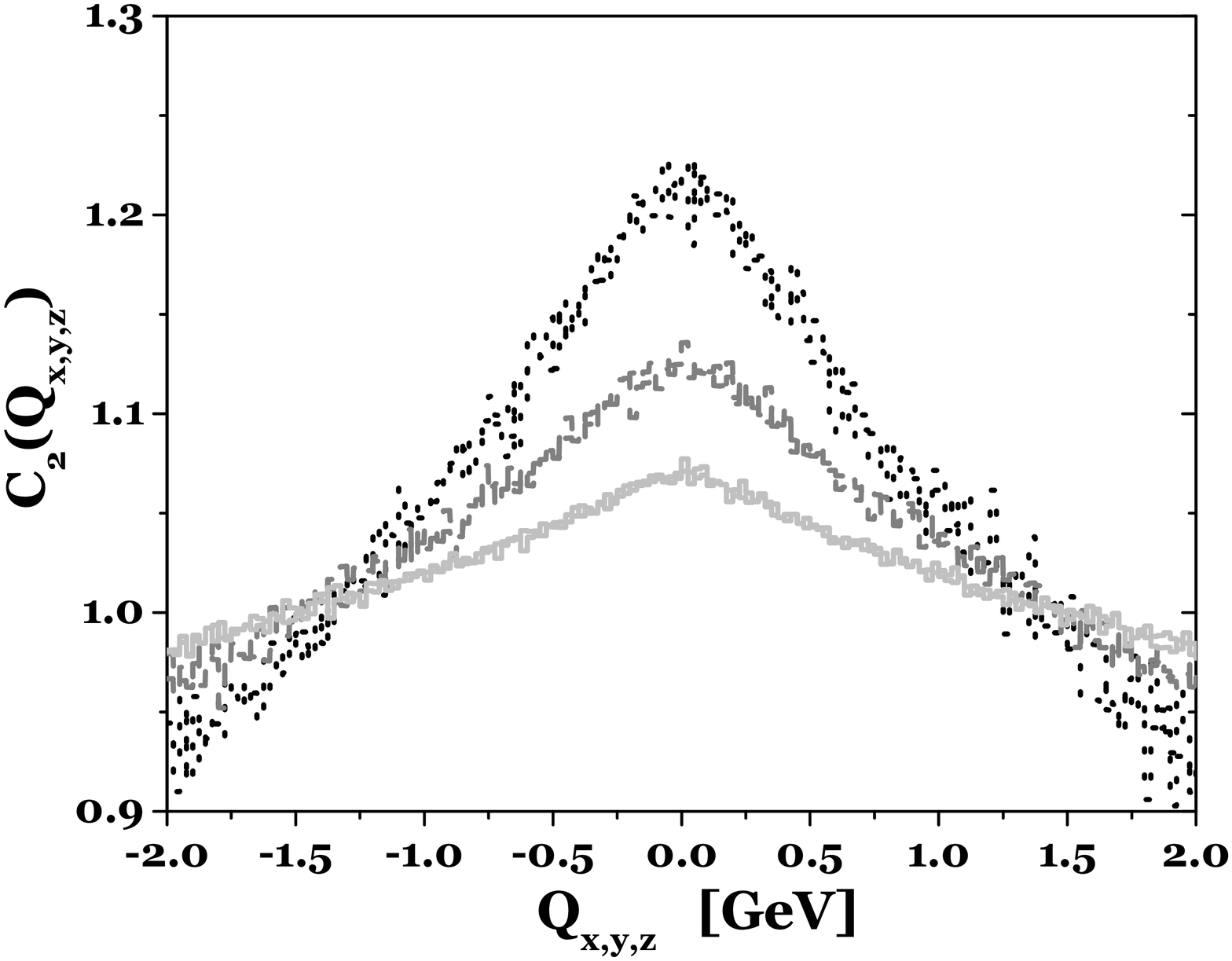}
\end{figure}
\begin{figure}
     \includegraphics[height=.25\textheight]{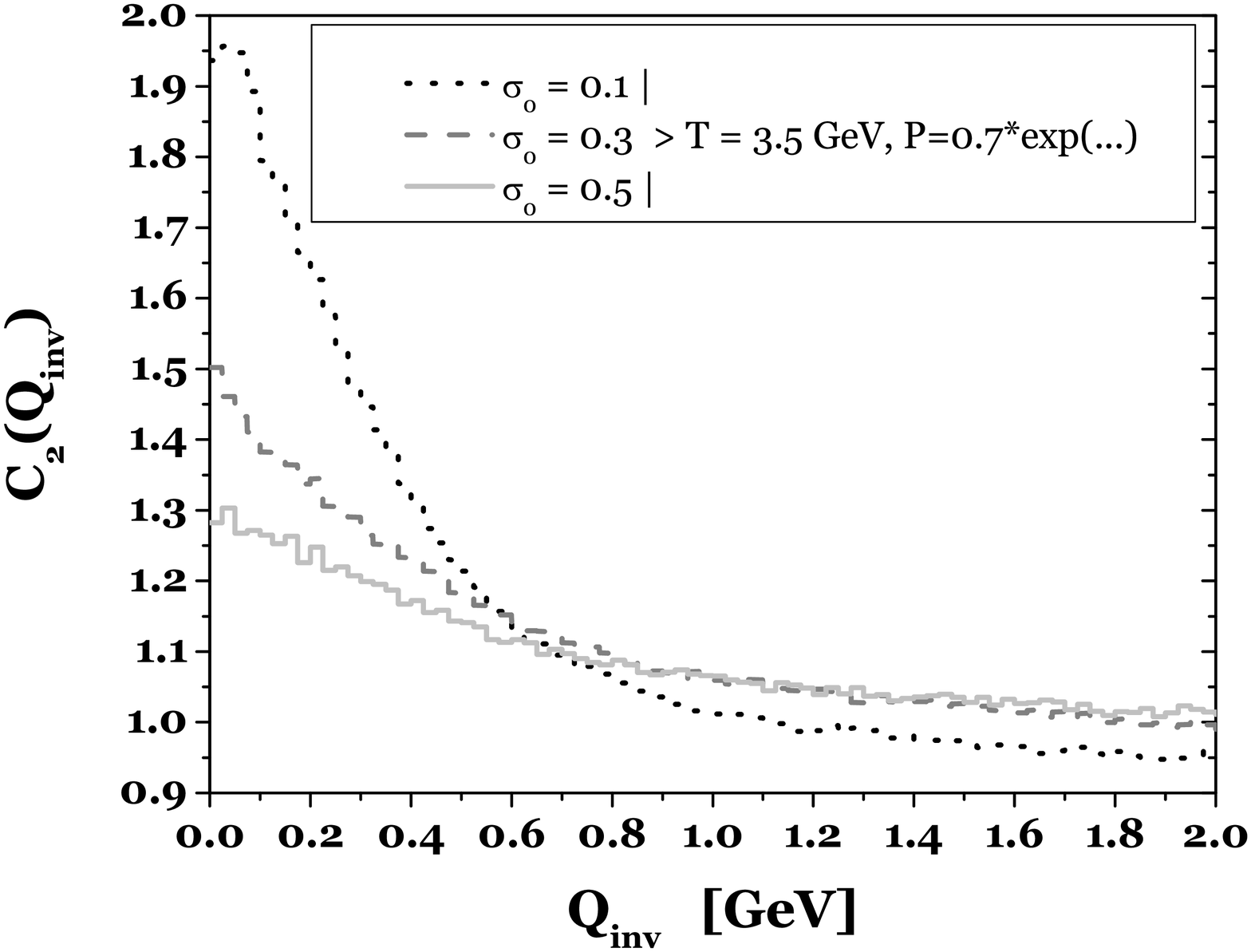}
  \hspace{3mm}
    \includegraphics[height=.25\textheight]{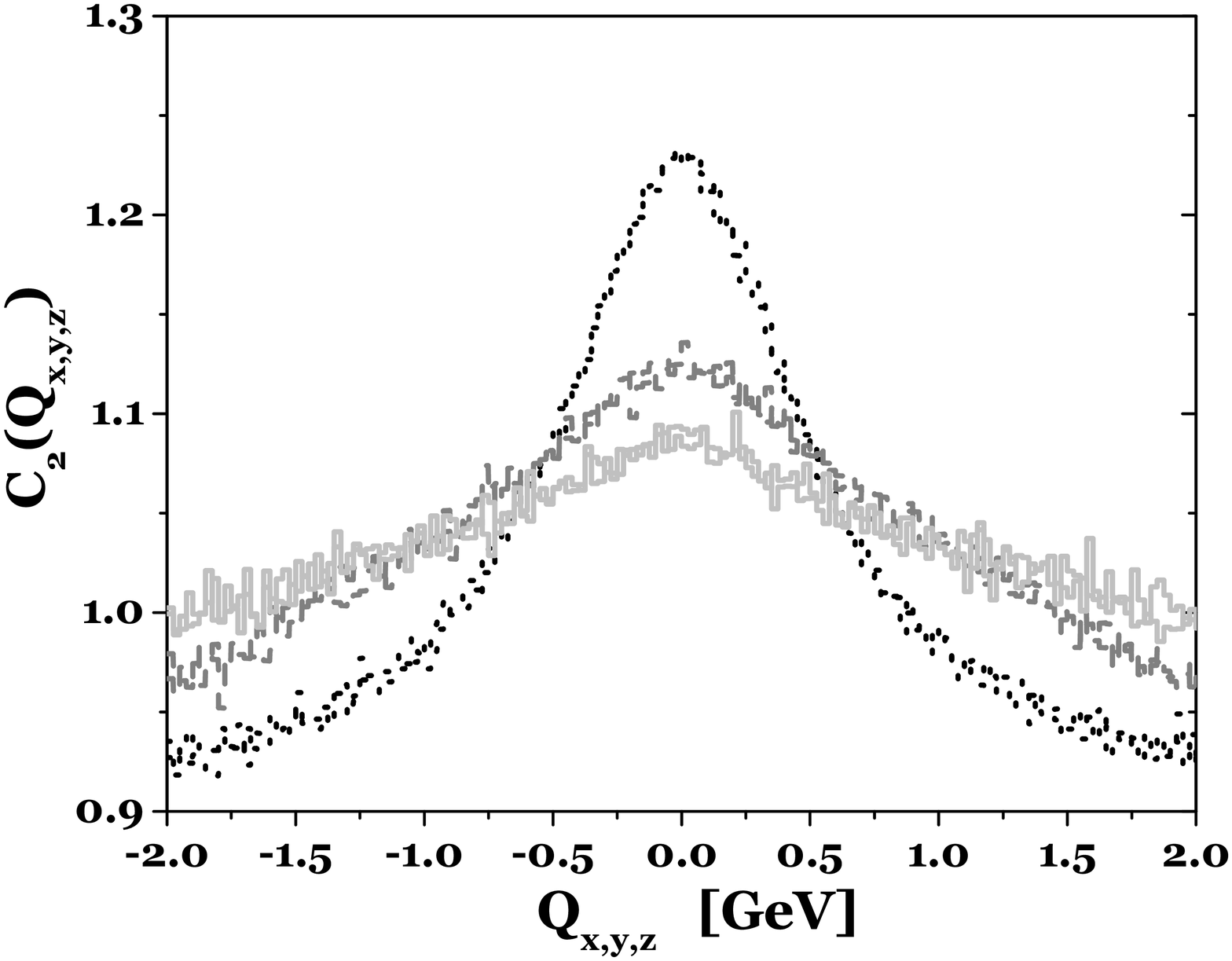}
  \caption{Example of results obtained with $3$-dimensional version of our model.
  Upper panels show results for different mass of the hadronizing source. Lower
  panel demonstrates dependence on the allowed energy spread in EEC.}
  \label{fig:Fig2}
\end{figure}


So far we are assuming direct pion production. However, the inclusion of
Coulomb and other final state interactions in our approach is
straightforward - one must simply change $\cos(\cdots)$ term arising from
plane wave approximation used to form obtained by some distorted wave
function. One can also easily include resonances and allow for finite
life time of the emitting source. Finally, so far, for simplicity reason,
only two particle symmetrization effects have been accounted for. Namely,
in a given EEC all particles are symmetrized with the particle number $1$
being its seed, they are not symmetrized between themselves. This seems
to be justified because majority of our EEC's contain only $1-2$ or $3$
particles. But to fully account for multiparticle effects one should
simply add other terms in addition to the $\cos(\dots)$ used above. This,
however, would result in dramatic increase of the calculational
time\footnote{In fact, this converges to the proposition presented long
time ago in \cite{ZAJC}, the only hope is that in our case symmetrization
is performed within given EEC and therefore the number of terms of
$\cos(\dots)$ type involved is rather limited, whereas in \cite{ZAJC} the
whole source had to be symmetrized at once resulting with number of terms
growing like $n!$ where $n$ is observed multiplicity.}. Nevertheless the
effect of including at least terms when symmetrization between, say,
particles $2$ and $3$ are added to the already present symmetrization
between $1$ and $2$ and $1$ and $3$, must be carefully investigated
before any final conclusion is to be reached.

\paragraph{Summary}

To summarize, we are proposing numerical scheme of modelling quantum
statistical phenomenon represented by BEC occurring in all hadronization
processes. Distinctive features of our scheme not present in other
propositions are:
\begin{itemize}
\item identical particles are emitted from EEC's and only these particles
are subjected to BEC;

\item inside each EEC particles are distributed according to the
geometrical (or Bose-Einstein) distribution;

\item altogether they show characteristic Bose-Einstein form of
distribution of energy.
\end{itemize}
As result we obtain a kind of {\it quantum clan model} with
Negative-Binomial-like multiplicity distributions and characteristic
shape of $C_2(Q_{inv})$ function \cite{QCM} (notice that we automatically
include in this way BEC to all orders given by the maximal number of
particles in a given EEC). To get also $C_2(Q_{x,y,z})$ one has to use
some additional space-time information and the characteristic
$1+\cos(\delta r\cdot\delta p)$ correlations between space-time and
energy-momenta induced by the symmetrization of the respective wave
functions. So far this is only a case study, we cannot yet offer any
attempt to compare it with experimental data. On the other hand our
approach offers new understanding of the way in which BEC are entering
hadronization process.

We shall close with remark that there are attempts in the literature to
model numerically BE condensation \cite{Condens} (or to use notion of BE
condensation in other branches of science as well \cite{ClassBE}) using
ideas of bunching of some quantities in the respective phase spaces.

\begin{theacknowledgments}
OU is grateful for support and for the warm hospitality extended to him
by organizers of the ISMD2005. Partial support of the Polish State
Committee for Scientific Research (KBN) (grant
621/E-78/SPUB/CERN/P-03/DZ4/99 (GW)) is acknowledged.
\end{theacknowledgments}

\end{document}